\documentstyle[prl,aps,twocolumn]{revtex}

\begin{document}
\title{Conditions for manipulation of a set of entangled pure states}
\author{Chuan-Wei Zhang, Chuan-Feng Li\thanks{%
Electronic address: cfli@ustc.edu.cn}, and Guang-Can Guo\thanks{%
Electronic address: gcguo@ustc.edu.cn}}
\address{Laboratory of Quantum Communication and Quantum Computation and Department\\
of Physics, \\
University of Science and Technology of China,\\
Hefei 230026, People's Republic of China}
\maketitle

\begin{abstract}
\baselineskip14ptWe derive a sufficient condition for a set of pure states,
each entangled in two remote $N$-dimensional systems, to be transformable to 
$k$-dimensional-subspace equivalent entangled states ($k\leq N$) by same
local operations and classical communication. If $k=N$, the condition is
also necessary. This condition reveals the function of the relative marginal
density operators of the entangled states in the entanglement manipulation
without sufficient information of the initial states.

\end{abstract}

\baselineskip14ptThe deep ways that quantum information differs from
classical information involve the properties, implications, and uses of {\it %
quantum entanglement }$[1]$. As a useful physical resource of quantum
information, entanglement plays a key role for quantum computation $[2]$,
quantum teleportation $[3]$, quantum superdense coding $[4]$ and certain
types of quantum cryptography $[5]$, etc. To accomplish these tasks,
transformation between the input entanglement we possess and the target
entanglement we require is necessary. Attempts $\left[ 6-18\right] $ have
been made to uncover the fundamental laws of the transformations under local
quantum operations and classical communication (LQCC), that is, the
different entangled parties may do whatever they wish to in their local
system, and may communicate classically, but they cannot use quantum
communication.

All previous entanglement manipulation protocols only consider a definite
entangled state shared by distant observers. However, quantum information
processing often has to work with insufficiently known initial states. It is
therefore important to understand which processes work without full
knowledge of initial states. In this letter, we address the question whether
it is possible to manipulate a set of entangled pure states only by one LQCC
protocol, just like quantum clone $[19-22]$. This problem is fundamentally
and also practically important. An example may be in the disentangled
process of quantum clone $\left[ 23-24\right] $. In deterministic
state-dependent cloning process $\left[ 21\right] $, although according to
Nielsen Theorem $\left[ 11\right] $ each of the two final states can be
transformed to the disentangled state by LQCC respectively, it is impossible
to separate the output by LQCC without knowing which one the initial state
is $\left[ 23\right] $. The same result also exists in probabilistic
telecloning process $\left[ 24\right] $. In Ref. $\left[ 18\right] $, we
showed that a local operation can enhance the entanglement of a set of
two-level entangled states simultaneously. In this letter, we investigate
the problem with some restrictions of the final states. The investigations
here are for the finite (nonasymptotic) case, from which asymptotic results
may be recovered by taking limits. The transformation process may be
probabilistic, but not approximate. To present our questions and results, we
first collect some useful {\bf Fact}s:

{\bf 1.} An arbitrary bipartite entangled pure state $\left| \Omega
\right\rangle $ that Alice and Bob share can be written as $\left[ 25\right] 
$ $\left| \Omega \right\rangle =\left( U_A\otimes U_B\right)
\sum\limits_{i=1}^N\sqrt{\lambda _i}\left| i_A\right\rangle \left|
i_B\right\rangle $, where $U_A\ $and $U_B$ are local unitary transformations
by Alice and Bob respectively, $\sum\limits_{i=1}^N\lambda _i=1$, and $%
\left\{ \left| i_A\right\rangle \right\} $ ($\left\{ \left| i_B\right\rangle
\right\} $) form an orthogonal basis for system $A$ ($B$). In this letter,
we take $N$ as the maximum dimensions of the subsystem. If $\left| \Omega
\right\rangle $ has $m$ no-zero eigenvalues, we call $\left| \Omega
\right\rangle $ as $m${\it -dimensional entangled state}. The $m$%
-dimensional maximally entangled state can be generally written as $\left|
\Phi \right\rangle =\left( U_A\otimes U_B\right) \frac 1{\sqrt{m}}%
\sum\limits_{i=1}^m\left| i_A\right\rangle \left| i_B\right\rangle $. The
marginal density operator for Alice's (Bob's) subsystem is defined as $\rho
_{A\left( B\right) }\left( \left| \Omega \right\rangle \right) =%
\mathop{\rm Tr}
_{B(A)}\left| \Omega \right\rangle \left\langle \Omega \right| $. Obviously $%
\lambda _i$ is the eigenvalue of $\rho _{A\left( B\right) }\left( \left|
\Omega \right\rangle \right) $. Furthermore, we denote $\left| \alpha
\right\rangle \sim \left| \beta \right\rangle $ if $\left| \alpha
\right\rangle $ and $\left| \beta \right\rangle $ are the same up to local
unitary operations by Alice and Bob. The Schmidt decomposition implies that $%
\left| \alpha \right\rangle \sim \left| \beta \right\rangle $ if and only if 
$\rho _A\left( \left| \alpha \right\rangle \right) $ and $\rho _A\left(
\left| \beta \right\rangle \right) $ have the same spectrum of eigenvalues.

{\bf 2.} Denote $Q$ as an index set. We call that a set of entangled states $%
\left\{ \left| \alpha _\ell \right\rangle ,\ell \in Q\right\} $ are $k${\it %
-dimensional-subspace equivalent} if and only if there exist no-zero
constant $C_{\alpha _\ell }$ and no-zero Schmidt coefficients $\mu _t\left(
\left| \alpha _\ell \right\rangle \right) $ ($1\leq t\leq k$) making $\mu
_t\left( \left| \alpha _{\ell _0}\right\rangle \right) =C_{\alpha _\ell }\mu
_t\left( \left| \alpha _\ell \right\rangle \right) $. Suppose $\left| \alpha
\right\rangle $ and $\left| \beta \right\rangle $ are $N$ and $N^{^{\prime
}} $ ($N^{^{\prime }}\leq N$) dimensional entangled states respectively, we
denote $\hat F_{A\left( B\right) }\left( \left| \alpha \right\rangle \left|
\beta \right\rangle \right) =\rho _{A\left( B\right) }^{-\frac 12}\left(
\left| \alpha \right\rangle \right) \rho _{A\left( B\right) }\left( \left|
\beta \right\rangle \right) \rho _{A\left( B\right) }^{-\frac 12}\left(
\left| \alpha \right\rangle \right) $ as the {\it relative marginal density
operator} of states $\left| \alpha \right\rangle $ and $\left| \beta
\right\rangle $ for Alice's (Bob's) subsystem. $\hat F_{A(B)}$ describes the
relation of the marginal density operators of the two states. For a set of
relative marginal density operators $\left\{ \hat F_{A\left( B\right) }^\ell
,\ell \in Q\right\} $, we denote that $\hat F_{A\left( B\right) }^\ell $ are 
{\it similar about }$I_k$ if and only if $\hat F_{A\left( B\right) }^\ell $
can be represented on the orthogonal basis $\left| i_A\right\rangle $ as 
\begin{equation}
\hat F_{A\left( B\right) }^\ell =V%
\mathop{\rm diag}
\left( s_\ell I_k,D_\ell \right) V,  \eqnum{1}
\end{equation}
where $s_\ell >0$, $I_k$ is the $k\times k$ unit matrix, $V$ is unitary and $%
D_\ell $ is a symmetric matrix.

{\bf 3.} Any operation $P$ Alice performs on the maximally entangled state $%
\frac 1{\sqrt{N}}\sum\limits_{k=1}^N\left| k_A\right\rangle \left|
k_B\right\rangle $ is equal to the transposed operation $P^{+}$ performed by
Bob $\left[ 26\right] $, that is, $\left( P\otimes I\right)
\sum\limits_{k=1}^N\left| k_A\right\rangle \left| k_B\right\rangle =\left(
I\otimes P^{+}\right) \sum\limits_{i=1}^N\left| k_A\right\rangle \left|
k_B\right\rangle $.

{\bf 4.} Given any pure bipartite state $\left| \Psi \right\rangle
_{AB}=\sum\limits_{i=1}^N\sqrt{\lambda _i}\left| i_A\right\rangle \left|
i_B\right\rangle $ shared by Alice and Bob and any complete set of
projection operators $\left\{ P_l^{Bob}\right\} $'s by Bob, there exists a
complete set of projection operators $\left\{ P_l^{Alice}\right\} $'s by
Alice and, for each outcome $l$, a direct product of local unitary
transformations $U_l^A\otimes U_l^B$ such that, for each $l$ $\left[
8\right] $%
\begin{equation}
\left( I\otimes P_l^{Bob}\right) \left| \Psi \right\rangle _{AB}=\left(
U_l^A\otimes U_l^B\right) \left( P_l^{Alice}\otimes I\right) \left| \Psi
\right\rangle _{AB}.  \eqnum{2}
\end{equation}

{\bf 5.} The most general scheme of entanglement manipulation of a bipartite
entangled pure state involves local operations of respective system and
two-way communication between Alice and Bob $\left[ 6\right] $. The local
operations can be represented as generalized measurements, described by
operators $A_k$ and $B_l$ on each system, satisfying the condition $%
\sum_kA_k^{+}A_k\leq I_N$ ($I_N-\sum_kA_k^{+}A_k$ is positive semidefinite)
and $\sum_lB_l^{+}B_l\leq I_N$, where $I_N$ is the unit operator of Alice's
or Bob's subsystem. The LQCC protocol we consider maps the initial state $%
\left| \phi \right\rangle \left\langle \phi \right| $ to the target state $%
\left[ 9\right] $, 
\begin{equation}
\left| \varphi \right\rangle \left\langle \varphi \right| =\frac{%
\sum_{kl}A_k\otimes B_l\left| \phi \right\rangle \left\langle \phi \right|
A_k^{+}\otimes B_l^{+}}{Tr\left( \sum_{kl}A_k\otimes B_l\left| \phi
\right\rangle \left\langle \phi \right| A_k^{+}\otimes B_l^{+}\right) }. 
\eqnum{3}
\end{equation}
The initial and final states are pure, it follows that 
\begin{equation}
A_k\otimes B_l\left| \phi \right\rangle =\sqrt{p_{kl}}\left| \varphi
\right\rangle ,  \eqnum{4}
\end{equation}
with non-negative success probability $p_{kl}$ satisfying $p_{kl}=Tr\left(
A_k\otimes B_l\left| \phi \right\rangle \left\langle \phi \right|
A_k^{+}\otimes B_l^{+}\right) $.

Suppose Alice and Bob share a pure bipartite $N$-dimensional entangled state 
$\left| \phi _1\right\rangle $ that they can convert to another entangled
pure state $\left| \varphi _1\right\rangle $ by a LQCC process with no-zero
probability $\left[ 12\right] $. Denote $S$ as an index set, our question is
what property characterizes the set of entangled pure state $\left\{ \left|
\phi _1\right\rangle ,\left| \phi _\nu \right\rangle ,\nu \in S\right\} $
that can be transformed to the final states $\left\{ \left| \varphi
_1\right\rangle ,\left| \varphi _\nu \right\rangle ,\nu \in S\right\} $ by
the same LQCC process if $\left| \varphi _\nu \right\rangle $ are $k$%
-dimensional-subspace equivalent to state $\left| \varphi _1\right\rangle $ (%
$k\leq N$). In this letter, we derive a sufficient condition for such
manipulation. If $k=N$, we show that the condition is also necessary.

{\bf Theorem 1: }A set of bipartite entangled pure states $\left\{ \left|
\varphi _1\right\rangle ,\left| \varphi _\nu \right\rangle ,\nu \in
S\right\} $ can be probabilistic transformed to $k$-dimensional-subspace
equivalent states by one LQCC protocol if the relative marginal density
operators of states $\left| \varphi _1\right\rangle $ and $\left| \varphi
_\nu \right\rangle $ are similar about $I_k$.

As a simple application of the result, suppose Alice and Bob each possess a
four-dimensional quantum system, with respectively orthonormal bases denoted
by $\left| 1\right\rangle $, $\left| 2\right\rangle $, $\left|
3\right\rangle $ and $\left| 4\right\rangle $. The initial entangled state
may be one of the following states 
\begin{eqnarray}
\left| \alpha \right\rangle &=&\sqrt{\frac 14}\left| 11\right\rangle +\sqrt{%
\frac 14}\left| 22\right\rangle +\sqrt{\frac 1{16}}\left| 33\right\rangle +%
\sqrt{\frac 7{16}}\left| 44\right\rangle ,  \eqnum{5} \\
\left| \beta \right\rangle &=&\sqrt{\frac 14}\left| 11\right\rangle +\sqrt{%
\frac 14}\left| 22\right\rangle +\sqrt{\frac 12}\left| 33\right\rangle . 
\nonumber
\end{eqnarray}
Obviously the relative marginal density operator $F_A\left( \left| \alpha
\right\rangle \left| \beta \right\rangle \right) =%
\mathop{\rm diag}
\left( 1,1,8,0\right) $ has two same eigenvalues. Alice can transform above
two states to $2$-dimensional maximally entangled state $\left| \Upsilon
\right\rangle =\sqrt{\frac 12}\left( \left| 11\right\rangle +\left|
22\right\rangle \right) $ with local generalized measurement $P_1=\left|
1\right\rangle \left\langle 1\right| +\left| 2\right\rangle \left\langle
2\right| $ satisfying $P_1^{+}P_1\leq I_4$.

{\bf Proof of Theorem 1:}

Generally, the states to be transformed can be represented as $\left| \phi
_1\right\rangle =\sum\limits_{i=1}^N\sqrt{\lambda _i}\left| i_A\right\rangle
\left| i_B\right\rangle $ and $\left| \phi _\nu \right\rangle =\left(
U_A^\nu \otimes U_B^\nu \right) \sum\limits_{i=1}^N\sqrt{\mu _i^\nu }\left|
i_A\right\rangle \left| i_B\right\rangle $ with $\lambda _i>0$. Suppose $%
\left| \phi _1\right\rangle $ is transformed to the state $\left| \varphi
_1\right\rangle =\sum\limits_{i=1}^N\sqrt{\gamma _i}\left| i_A\right\rangle
\left| i_B\right\rangle $ by a LQCC process, the same LQCC should transform
state $\left| \phi _\nu \right\rangle $ to state $\left| \varphi _\nu
\right\rangle $ that has Schmidt coefficients $\eta _i^\nu =c_\nu \gamma _i$%
, $i=1,2,...,k$, where $c_\nu $ is a no-zero real number.

Denote $\lambda =%
\mathop{\rm diag}
\left( \lambda _1,\lambda _2,...,\lambda _N\right) $, and $\mu ^\nu $, $%
\gamma $, $\eta ^\nu $ are of similar definitions. Obviously $\rho _A\left(
\left| \phi _1\right\rangle \right) =\lambda $, $\rho _A\left( \left| \phi
_\nu \right\rangle \right) =U_A^\nu \mu ^\nu \left( U_A^\nu \right) ^{+}$.
Since the relative marginal density operators $\left\{ \hat F_A^\nu \left(
\left| \varphi _1\right\rangle \left| \varphi _\nu \right\rangle \right)
,\nu \in S\right\} $ are similar about $I_k$, applying Fact {\bf 2}, we
obtain 
\begin{equation}
\lambda ^{-\frac 12}U_A^\nu \sqrt{\mu ^\nu }=V\left( 
\begin{array}{cc}
\sqrt{s_\nu }I_k & 0 \\ 
0 & \sqrt{D_\nu }
\end{array}
\right) G_\nu ,  \eqnum{6}
\end{equation}
where $G_\nu $ is a unitary matrix. Suppose $P_l=\sqrt{\varepsilon _l}\sqrt{%
\gamma }V^{+}\sqrt{\lambda ^{-1}}$. Since $P_l^{+}P_l=\varepsilon _l\sqrt{%
\lambda ^{-1}}V\gamma V^{+}\sqrt{\lambda ^{-1}}$, suitable choice of $%
\varepsilon _l$ can make $P_l^{+}P_l\leq I_N$, which means $P_l$ is a
generalized measurement (Fact {\bf 5}) independent of the initial states $%
\left| \phi _\nu \right\rangle $. According to Fact {\bf 3}, $P_l$ acts on
the states $\left| \phi _1\right\rangle $ and $\left| \phi _\nu
\right\rangle $ as follows:
\begin{eqnarray}
\left( P_l\otimes I\right) \left| \phi _1\right\rangle  &=&\sqrt{\varepsilon
_l}\left( I\otimes V\right) \sum_{i=1}^N\sqrt{\gamma _i}\left|
i_A\right\rangle \left| i_B\right\rangle   \eqnum{7} \\
&=&\sqrt{\varepsilon _l}\left( I\otimes V\right) \left| \varphi
_1\right\rangle ,  \nonumber
\end{eqnarray}
\begin{eqnarray}
&&\left( P_l\otimes I\right) \left| \phi _\nu \right\rangle   \eqnum{8} \\
&=&\sqrt{\varepsilon _ls_\nu }\left( I\otimes U_B^\nu G_\nu ^{+}V\right)
\left( I\otimes H_\nu \right) \sum_{i=1}^N\sqrt{\gamma _i}\left|
i_A\right\rangle \left| i_B\right\rangle   \nonumber \\
&=&\sqrt{\varepsilon _ls_\nu }\left( I\otimes U_B^\nu G_\nu ^{+}V\right)
\left| \varphi _2^{^{\prime }}\right\rangle ,  \nonumber
\end{eqnarray}
where the corresponding matrix of $H_\nu =%
\mathop{\rm diag}
\left( I_k,s_\nu ^{-1}\sqrt{D_\nu }\right) $, $\left| \varphi _\nu
^{^{\prime }}\right\rangle =\left( I\otimes H_\nu \right) $ $\left.
\sum\limits_{i=1}^N\sqrt{\gamma _i}\left| i_A\right\rangle \left|
i_B\right\rangle \right. $. Denote the normalized states of $\left| \varphi
_\nu ^{^{\prime }}\right\rangle $ as $\left| \varphi _\nu \right\rangle $,
they are $k$-dimensional-subspace equivalent to $\left| \varphi
_1\right\rangle $. Thus we finish the {\bf proof of Theorem 1}.

If the final states are $N$-dimensional-subspace equivalent, the above
sufficient condition can be expressed in a more simple form with clear
physical meaning. In fact, since $%
\mathop{\rm Tr}
\rho _A\left( \phi _1\right) =%
\mathop{\rm Tr}
\rho _A\left( \left| \phi _\nu \right\rangle \right) =1$, the above
sufficient condition means that the relative marginal density operators $%
F_A^\nu \left( \left| \varphi _1\right\rangle \left| \varphi _\nu
\right\rangle \right) =I$, i.e., the marginal density operators $\rho
_A\left( \phi _1\right) =\rho _A\left( \left| \phi _\nu \right\rangle
\right) $ and $\left| \phi _\nu \right\rangle $ are also $N$-dimensional
entangled pure states. In this case, such condition is also necessary.

{\bf Theorem 2}: A set of $N$-dimensional entangled pure states $\left\{
\left| \phi _1\right\rangle ,\left| \phi _\nu \right\rangle ,\nu \in
S\right\} $ can be probabilistic transformed to $N$-dimensional-subspace
equivalent states $\left\{ \left| \varphi _1\right\rangle ,\left| \varphi
_\nu \right\rangle ,\nu \in S\right\} $ by same LQCC protocol if and only if
they share same marginal density operators for Alice's or Bob's subsystem.

{\bf Proof of Theorem 2:}

We need only prove the necessity. Consider that a generalized measurement $%
A_k\otimes B_l$ can transform the $N$-dimensional states $\left| \phi
_1\right\rangle $ and $\left| \phi _\nu \right\rangle $ to $N$%
-dimensional-subspace equivalent states{\it \ }$\left| \varphi
_1\right\rangle $ and $\left| \varphi _\nu \right\rangle $. Obviously the
Schmidt coefficients of states $\left| \varphi _1\right\rangle $ and $\left|
\varphi _\nu \right\rangle $ are greater than zero and $\left| \varphi
_1\right\rangle $ $\sim \left| \varphi _\nu \right\rangle $. We first prove
the necessity in the condition that only one side generalized measurement is
performed. The one-side generalized measurement acts on the initial states
as follows:
\begin{eqnarray}
&&\left( P_l\otimes I\right) \sum\limits_{i=1}^N\sqrt{\lambda _i}\left|
i_A\right\rangle \left| i_B\right\rangle   \eqnum{9} \\
&=&\sqrt{\varsigma }\left( E_1\otimes F_1\right) \sum_{i=1}^N\sqrt{\kappa _i}%
\left| i_A\right\rangle \left| i_B\right\rangle ,  \nonumber \\
&&\left( P_lU_A^\nu \otimes U_B^\nu \right) \sum\limits_{i=1}^N\sqrt{\mu
_i^\nu }\left| i_A\right\rangle \left| i_B\right\rangle   \nonumber \\
&=&\sqrt{\tau _\nu }\left( E_\nu \otimes F_\nu \right) \sum_{i=1}^N\sqrt{%
\kappa _i}\left| i_A\right\rangle \left| i_B\right\rangle ,  \nonumber
\end{eqnarray}
where $E_1\otimes F_1$ and $E_\nu \otimes F_\nu $ are local unitary
operations, $\varsigma $ and $\tau _\nu $ are the probabilities of success, $%
\kappa =%
\mathop{\rm diag}
\left( \kappa _1,\kappa _2,...,\kappa _N\right) $ is the eigenvalue matrix
of the final states. Since the final states are $N$-dimensional-subspace
equivalent, they must be $N$-dimensional entangled states (Fact {\bf 2}) and 
$\kappa _i>0$ for $i=1,2,...,N$. According to Fact {\bf 3}, the above two
equations can be represented with matrices as 
\begin{eqnarray}
P_l\sqrt{\lambda } &=&\sqrt{\varsigma }E_1\sqrt{\kappa }F_1^{+},  \eqnum{10}
\\
P_lU_A^\nu \sqrt{\mu ^\nu }\left( U_B^\nu \right) ^{+} &=&\sqrt{\tau _\nu }%
E_\nu \sqrt{\kappa }F_\nu ^{+}.  \nonumber
\end{eqnarray}
Substituting $P_l\sqrt{\lambda }$ of the first equation into the second, we
obtain 
\begin{equation}
T_\nu ^{+}\kappa T_\nu =\kappa ,  \eqnum{11}
\end{equation}
where $T_\nu =\sqrt{\frac \varsigma {\tau _\nu }}F_1^{+}\sqrt{\lambda ^{-1}}%
U_A^\nu \sqrt{\mu ^\nu }\left( U_B^\nu \right) ^{+}F_\nu $. Since $\kappa
_i>0$, Eq. (11) means that $T_\nu $ is unitary, it follows $\frac \varsigma {%
\tau _\nu }U_A^\nu \mu ^\nu \left( U_A^\nu \right) ^{+}=\lambda $. Since $%
\sum_i\lambda _i=1$, $\sum_i\mu _i^\nu =1$, we get $\tau _\nu =\varsigma $, $%
\mu ^\nu =\lambda $, and 
\begin{equation}
\rho _A\left( \left| \phi _1\right\rangle \right) =\lambda =U_A^\nu \mu ^\nu
\left( U_A^\nu \right) ^{+}=\rho _A\left( \left| \phi _\nu \right\rangle
\right) .  \eqnum{12}
\end{equation}

Now we consider that a two-side generalized measurement $A_k\otimes B_l$
transforms the input states $\left| \phi _1\right\rangle $ and $\left| \phi
_\nu \right\rangle $ to $N$-dimensional-subspace equivalent states. With
Fact {\bf 4}, we get
\begin{eqnarray}
&&A_k\otimes B_l\left| \phi _1\right\rangle   \eqnum{13} \\
&=&\left( A_kV_l^AB_l\otimes V_l^B\right) \left| \phi _1\right\rangle , 
\nonumber \\
&&\left( A_k\otimes B_l\right) \left| \phi _\nu \right\rangle   \nonumber \\
&=&\left( A_kU_A^\nu H_l^{\nu A}B_lU_B^\nu \otimes H_l^{\nu B}\right)
\sum\limits_{i=1}^N\sqrt{\mu _i^\nu }\left| i_A\right\rangle \left|
i_B\right\rangle ,  \nonumber
\end{eqnarray}
where $V_l^A$, $V_l^B$, $H_l^{\nu A}$ and $H_l^{\nu B}$ are local unitary
operations. The above two equations means that one-side generalized
measurement $A_k\otimes I$ can transform the initial states $\left(
V_l^AB_l\otimes I\right) \sum\limits_{i=1}^N\sqrt{\lambda _i}\left|
i_A\right\rangle \left| i_B\right\rangle $ and $\left( U_A^\nu H_l^{\nu
A}B_lU_B^\nu \otimes I\right) \sum\limits_{i=1}^N\sqrt{\mu _i^\nu }\left|
i_A\right\rangle \left| i_B\right\rangle $ to $N$-dimensional-subspace
equivalent states. Therefore $\left( B_l\otimes I\right) \sum\limits_{i=1}^N%
\sqrt{\lambda _i}\left| i_A\right\rangle \left| i_B\right\rangle $ and $%
\left( B_lU_B^\nu \otimes I\right) \sum\limits_{i=1}^N\sqrt{\mu _i^\nu }%
\left| i_A\right\rangle \left| i_B\right\rangle $ must also be $N$%
-dimensional-subspace equivalent states, which means the marginal density
operators for Bob's side of the input states must satisfy $\rho _B\left(
\left| \phi _1\right\rangle \right) =\rho _B\left( \left| \phi _\nu
\right\rangle \right) $. So one of the two subsystems of the initial states
must have same marginal density operators.

So far we have proven {\bf Theorem 1 }and {\bf Theorem 2}. In {\bf Theorem 1}%
, we give a sufficient condition for that a set of entangled pure states can
be probabilistic transformed to $k$-dimensional-subspace equivalent states
by same LQCC protocol. We conjecture that this condition is also necessary.
In fact, it is true if the generalized measurement is restricted in one
side. In this case, the eigenvalue matrix of the final states in Eq. (10) is
substituted by $\kappa ^\nu $. $\kappa ^\nu $ should have at least $k_\nu $ (%
$k_\nu \geq k$) no-zero eigenvalues $\kappa _i^\nu =d_\nu \kappa _i$, $1\leq
i\leq k_\nu $, where $d_\nu $ is a constant dependent on $\nu $. Eq. (11)
should be rewritten as 
\begin{equation}
T_\nu ^{+}\kappa T_\nu =\kappa ^{^\nu },  \eqnum{14}
\end{equation}
where $T_\nu $ is the same as that in Eq. (11). Eq. (14) means $T_\nu $ can
be represented as 
\begin{equation}
T_\nu =\left( 
\begin{array}{cc}
\sqrt{d_\nu }M_{k_\nu }^\nu  & 0 \\ 
0 & R_\nu 
\end{array}
\right) ,  \eqnum{15}
\end{equation}
where $M_{k_\nu }^\nu $ is a $k_\nu \times k_\nu $ unitary matrix and $R_\nu 
$ may be any possible matrix. The unitarity of $M_{k_\nu }^\nu $ yields
\begin{eqnarray}
\hat F_A\left( \left| \phi _1\right\rangle \left| \phi _\nu \right\rangle
\right)  &=&\lambda ^{-\frac 12}U_A^\nu \mu ^\nu \left( U_A^\nu \right)
^{+}\lambda ^{-\frac 12}  \eqnum{16} \\
&=&\frac{d_\nu \tau _\nu }\varsigma F_1\left( 
\begin{array}{cc}
I_{k_\nu } & 0 \\ 
0 & \frac 1dR_\nu R_\nu ^{+}
\end{array}
\right) F_1^{+}.  \nonumber
\end{eqnarray}
Since $k_\nu \geq k$ and $F_1$ is independent of index $\nu $, Eq. (16)
means that the relative marginal density operators $\hat F_A\left( \left|
\phi _1\right\rangle \left| \phi _\nu \right\rangle \right) $ of the initial
states $\left| \phi _1\right\rangle $ and $\left| \phi _\nu \right\rangle $
are similar about $I_k${\it . }

{\bf Theorem 2} shows that the sufficient condition in {\bf Theorem 1} is
also necessary in a special case. The result means that Alice (Bob) cannot
probabilistically transform $N$-dimensional entangled states that are
different in her (his) local observation to $N$-dimensional-subspace
equivalent states. Generally the ordered Schmidt coefficients of the states
to be transformed must be the same, but these states need not be the same,
there exist unitary transformations on both Alice's and Bob's sides. While
arbitrary on Bob's (Alice's) side, the unitary operators on Alice's (Bob's)
side must preserve the density matrix $\rho _A$ ($\rho _B$), which means
that only when there exist some coefficients satisfying $\lambda _i=\lambda
_{i+1}$, the unitary operators $U_A^\nu $ ($U_B^\nu $) can be non-unit.

The above results can be directly applied to {\it concentration of
entanglement} $\left[ 6,8\right] $, that is, transforming partial
entanglement to maximally entanglement. {\bf Theorem 1} gives a sufficient
condition for the concentration of a set of partial entanglement to the
maximally entanglement (not necessary $N$-dimensional), while {\bf Theorem 2}
shows that Alice (Bob) can probabilistic concentrate several different $N$%
-dimensional partial entangled states to $N$-dimensional maximally entangled
states by same LQCC process if and only if the marginal density operators of
these states are the same for her or his subsystem. In the proof of {\bf %
Theorem 2} we also showed the following important result:

{\bf Proposition 1}: Different $N$-dimensional entangled states cannot be
transformed to one $N$-dimensional entangled state by same LQCC protocol on
individual pairs.

However, such result does not prohibit us from transforming different
entangled states to one of lower dimension. An example is the states $\left|
\alpha \right\rangle $ and $\left| \beta \right\rangle $ in Eqs. (5).

While one can always, with finite probability, bring an individual entangled
pure state to a maximally entangled state using only LQCC $\left[ 8\right] $%
, Linden {\it et al.} $[9]$ have shown that it is impossible to purify a
two-level mixed state to a maximally entangled state by any combination of
LQCC acting on individual pairs. In this letter we generalize it to $N$%
-level mixed state as

{\bf Theorem 3 }: It is impossible to purify a $N$-dimensional mixed state
to a $N$-dimensional maximally entangled state by LQCC on individual pairs.

{\bf Proof of Theorem 3:}

Consider a given mixed state $\rho $, generally we can use the spectral
decomposition $\left[ 25\right] $ of the state $\rho =\sum_ip_i\left| \psi
_i\right\rangle \left\langle \psi _i\right| $. {\bf Proposition 1} indicates
that different decomposition terms $\left| \psi _i\right\rangle $ of the
mixed state $\rho $ can never be transformed to one $N$-dimensional pure
state by same LQCC, which means $\rho $ cannot be concentrated into a $N$%
-dimensional maximally entangled pure state by LQCC on individual pairs.
This result is surprising because we expect entanglement to be a property of
each pair individually rather than a global property of many pairs.

However, {\bf Theorem 3} does not mean that we cannot obtain lower
dimensional entangled pure state from a mixed state by LQCC. For example, we
can concentrate the mixed state $\rho =\frac 14\left| \alpha \right\rangle
\left\langle \alpha \right| +\frac 34\left| \beta \right\rangle \left\langle
\beta \right| $ to the maximally entangled pure state $\left| \Upsilon
\right\rangle =\sqrt{\frac 12}\left( \left| 11\right\rangle +\left|
22\right\rangle \right) $, where $\left| \alpha \right\rangle $ and $\left|
\beta \right\rangle $ are the states in Eqs. (5).

Another interesting application may be probabilistic quantum superdense
coding. Suppose Bob has four choices to perform $U_B$ i.e. $\left\{ I,\sigma
_x,i\sigma _y,\sigma _z\right\} $ on the initial possessed partial entangled
states, just like that in Ref. $[4]$. Alice still can transform the partial
entangled state to the maximally entangled state with no-zero probability,
although she does not know which $U_B$ Bob performs. Bob sends his particle
to Alice after he has performed $U_B$. Alice's task is then to identify the
four Bell states and obtain the information.

The further application of these results need to be explored. Similar to
quantum cloning process, although we lack sufficient information about the
initial states, we still can make operations on them and extract information
at the end. The indefinite initial entanglement may contain quantum
information and our results may be useful in quantum cryptography and
quantum communication.

In summary, we have shown that a set of entangled pure states $\left\{
\left| \varphi _1\right\rangle ,\left| \varphi _\nu \right\rangle ,\nu \in
S\right\} $ can be probabilistic transformed to $k$-dimensional-subspace
equivalent states by same LQCC protocol if the relative marginal density
operators $\hat F_{A(B)}\left( \left| \varphi _1\right\rangle \left| \varphi
_\nu \right\rangle \right) $ are similar about $I_k$. In the case of that
the final states are $N$-dimensional-subspace equivalent, the condition can
be expressed as that the input states must share the same marginal density
operators for Alice's or Bob's subsystem and it is both sufficient and
necessary. As the application, we showed that it is impossible to purify a
mixed state to a maximally entangled state of same dimension by LQCC on
individual pairs and presented the probabilistic superdense coding.

{\bf Acknowledgment}: This work was supported by the National Natural
Science Foundation of China.

\end{document}